\documentclass[12pt]{JHEP3}
\preprint{  }
\usepackage{epsfig}
\usepackage{amsmath}
\usepackage{graphics}
\usepackage{graphicx}
\usepackage{dcolumn}
\usepackage{amssymb}
\usepackage{amsthm}
\usepackage{amsfonts}
\usepackage{subfigure}

\title{Generalized thermodynamic identity and new Maxwell's law for charged AdS black hole}
\author{Zixu Zhao and Jiliang Jing\footnote{Corresponding author. Email:
jljing@hunnu.edu.cn}
\\ Department of Physics, Key Laboratory of Low Dimensional Quantum Structures
and Quantum Control of Ministry of Education, and Synergetic Innovation Center
for Quantum Effects
and Applications, Hunan Normal University, Changsha, Hunan 410081, P. R. China.
}

\abstract{
We study the thermodynamic properties of the RN-AdS black hole in full
phase space and propose a generalized thermodynamic identity. As an example, we
use it to find relations of thermodynamical coefficients  between
the grand canonical and canonical ensembles. We also show,  for the
first order phase transition, that the usual Maxwell's equal area law
should be extended to a new form for the RN-AdS black hole.
}

\begin{document}

\section{Introduction}

Black holes are considered as thermodynamic systems \cite{Bekenstein,Hawking}, and their thermodynamic behaviors have been studied extensively \cite{Davies,Hawking2,Davies2,Hawking3,Carlip,Carter,Doneva}. The laws and behaviors of traditional thermodynamics have been expected to be reproduced in black hole systems although black holes have some unusual properties.
Chamblin \emph{et al.} studied the phase structures of the charged black hole systems and found that the structures isomorphic to the Van der Waals-Maxwell liquid-gas system in the canonical ensemble \cite{Chamblin}. Furthermore, they refined the phase diagrams in the canonical and grand canonical ensembles \cite{Chamblin2} and explored the physics in more detail because there exist striking similarity between the thermodynamic phase structure of these black holes and that of the Van der Waals-Maxwell liquid-gas system in the canonical ensemble. Later, Kubiz$\check{n}\acute{a}$k and Mann studied the $P-V$ behaviors of a charged AdS black hole \cite{KM}, which based on the great interest to include the variation of the cosmological constant in the first law of black hole thermodynamics \cite{Caldarelli,Kastor,Dolan,Dolan2,Dolan3,C,L}. On the other hand, the phase transition of thermodynamic systems has been investigated. After the discovery of a $\lambda$-shaped ``jump" discontinuity in the curve giving the temperature dependence of the specific heat of helium at a critical value, Ehrenfest \cite{Ehrenfest} introduced a classification of phase transitions on the basis of the behavior of the Gibbs free energy. There exist the usual Ehrenfest equations on the phase boundary for the second order phase transitions.  Recently, Ehrenfest equations have been studied in black hole thermodynamics \cite{Banerjee,Banerjee2,Banerjee3,Mo,Mo4}.

Since the usual thermodynamic behaviors and laws governing their application form a simple and elegant scheme, thus people would be inclined to think could not be modified. For examples, the usual Ehrenfest equations, thermodynamic identity and Maxwell's equal area law have been directly used to study black hole systems. Nevertheless it has been found possible to set up a new scheme, which is more suitable for the description of complex thermodynamic systems in full phase space. The RN-AdS black hole has three pairs of intensive/extensive variables temperature/entropy $(T,S)$, pressure/volume $(P,V)$ and electric potential/charge $(\Phi,Q)$, and its thermodynamical coefficients attain infinite values on the phase boundary, therefore we have proposed generalized Ehrenfest equations for the second order phase transition\cite{JHEP11}. Furthermore, we believe that some other laws and behaviors of traditional thermodynamics  should also be extended in black hole systems.

It is well-known that thermodynamic identity plays an important role in thermodynamics. However, the usual thermodynamic identity is suitable for the description of a thermodynamic system with two pairs of intensive/extensive variables. If we directly apply it to a system for three (or more) pairs  of intensive/extensive variables, we will get incomplete results. Therefore, we think that the usual thermodynamic identity should be extended for the RN-AdS black hole.  By using the new thermodynamic identity obtained in this paper we can find relations of thermodynamical coefficients  between the grand canonical and canonical ensembles.

On the other hand, in the first order phase transition there exist Maxwell's equal area law for $T<T_c$ in traditional thermodynamics in which the ``oscillating" part of the isotherm is replaced by an isobar so that the theoretical curves agree with actual measurements. We note that, a thermodynamic systems with multiple pairs of intensive/extensive variables would undergo a complicated phase transition process and we should not consider it as the simple usual thermodynamic process although some theoretical curves are similarity. An external disturbances may modify the actual evolution for the first phase transition. Therefore we cannot count on the usual Maxwell' s equal area law to save this accordance. In order to ensure the theoretical curves agree with actual evolution curve, in this paper we find another law --- new Maxwell's law,  which  describes the actual evolution of the first phase transition for a system with three pairs  of intensive/extensive variables.

The organization of the work is as follows. In sec 2, we present a generalized thermodynamic identity for  RN-AdS black hole, and we find the relations of thermodynamical coefficients in RN-AdS black hole between two ensembles  by using the identity. In sec 3 we present a new Maxwell's law for RN-AdS black hole. We will conclude in sec 4 of our main results.

\section{Generalized thermodynamic identity in RN-AdS black hole and its applacations}

We all know that for a thermodynamic system with two pairs of intensive/extensive variables $(T,S)$ and $(P,V)$, by selecting three variables from the system, such as $P,V,T$, one can obtain the equation of state $f(V,P,T)=0$, which gives thermodynamic identity
\begin{equation}\label{id}
\left(\frac{\partial V}{\partial P}\right)_{T}\left(\frac{\partial P}{\partial T}\right)_{V}\left(\frac{\partial T}{\partial V}\right)_{P}=-1.
\end{equation}

It should be noted that the description of $f(V,P,T)=0$ is incomplete for RN-AdS black hole because this black hole has three pairs of intensive/extensive variables $(T,S)$, $(P,V)$ and $(\Phi,Q)$. That is to say, for a thermodynamic system with three pairs (or more) of intensive/extensive variables, the thermodynamic identity should be extended.

\subsection{Generalized thermodynamic identity in RN-AdS black hole}

The equation of state for a thermodynamic system with three pairs of intensive/extensive variables can be expressed as
\begin{equation}\label{state}
f(x_1,x_2,x_3,y_i)=0,
\end{equation}
where $x_1, x_2$ and $ x_3$ is selected from every pair of intensive/extensive variables respectively, and $y_i$ take a conjugate variable for $x_1, x_2$ or $ x_3$.  We therefore have
\begin{eqnarray}\label{idNew}
\left(\frac{\partial x_1}{\partial x_2}\right)_{x_3,y_i}\left(\frac{\partial x_2}{\partial x_3}\right)_{y_i,x_1}
\left(\frac{\partial x_3}{\partial y_i}\right)_{x_1,x_2}\left(\frac{\partial y_i}{\partial x_1}\right)_{x_2,x_3}=1.
\end{eqnarray}

Now we apply the formula (\ref{idNew}) to the RN-AdS black hole with
the temperature, entropy and electric potential
\begin{eqnarray}
&&T=\left(1+{3r_+^2}/{l^2}-{Q^2}/{r_+^2}\right)/{4\pi r_+},\nonumber \\
&&S=\pi r_+^2 ,\nonumber \\ && \Phi={Q}/{r_+},\label{TSPhi}
\end{eqnarray}
where $l$ is the AdS radius, $Q$ is the electric charge,  and $r_+$ is the radius of the event horizon.
By using the suggestion \cite{Kastor,Dolan} that
\begin{eqnarray}
P={3}/{8\pi l^2} ,  \ \ \ \ \ \ V={4\pi r_+^3}/{3},
\end{eqnarray}
the first law of the black hole thermodynamics \cite{KM} and the corresponding Smarr relation \cite{Kastor} can be expressed as
\begin{eqnarray}
&&dM=TdS+\Phi dQ+VdP,\nonumber \\
&&M=2(TS-PV)+\Phi Q.
\end{eqnarray}

From the equation of state (\ref{state}) we know that there exist some different equations of state for RN-AdS black hole. For example, we can take the equation of state as $f_1(S,T,P,\Phi)=0$, then we get a thermodynamic identity
\begin{equation}
\left(\frac{\partial S}{\partial T}\right)_{P,\Phi}\left(\frac{\partial T}{\partial P}\right)_{\Phi,S}\left(\frac{\partial P}{\partial \Phi}\right)_{S,T}\left(\frac{\partial \Phi}{\partial S}\right)_{T,P}=1.\label{relation1}
\end{equation}
On the other hand, we can also take the equation of state as $f_2(S,T,P,Q)=0$ and  get another  thermodynamic identity\begin{equation}
\left(\frac{\partial S}{\partial T}\right)_{P,Q}\left(\frac{\partial T}{\partial P}\right)_{Q,S}\left(\frac{\partial P}{\partial Q}\right)_{S,T}\left(\frac{\partial Q}{\partial S}\right)_{T,P}=1. \label{relation2}
\end{equation}
In the following we will use the identities (\ref{relation1}) and (\ref{relation2}) to study the relations of thermodynamical coefficients  of RN-AdS black hole between grand canonical and canonical ensembles.

\subsection{Relations of thermodynamical coefficients
between grand canonical and canonical ensembles}

There exist some different behaviors for the RN-AdS black hole in the grand canonical and canonical ensembles. For a typical example, in the grand canonical ensembles, the specific heat capacity is given by \cite{Banerjee3}
\begin{equation}
C_{P,\Phi}=2S\frac{8PS^2-S-\pi Q^2}{8PS^2-S+\pi Q^2},
\end{equation}
which is different from the specific heat capacity
\begin{equation}
C_{P,Q}=2S\frac{8PS^2-S-\pi Q^2}{8PS^2-S+3\pi Q^2},
\end{equation}
obtained in the canonical ensembles \cite{KM}. We would like to find the relation between $C_{P,\Phi}$ and $C_{P,Q}$ by using the new thermodynamic identity for the RN-AdS black hole.

From Eqs. (\ref{relation1}) and  (\ref{relation2}),  we find
\begin{eqnarray}
C_{P,Q}=C_{P,\Phi}\left(\frac{\partial Q}{\partial \Phi}\right)_{S,T}\left(\frac{\partial \Phi}{\partial Q}\right)_{T,P}\left(\frac{\partial T}{\partial P}\right)_{\Phi,S}\left(\frac{\partial P}{\partial T}\right)_{Q,S}.\nonumber \\
\end{eqnarray}

Noting that
\begin{eqnarray}
&&\left(\frac{\partial Q}{\partial \Phi}\right)_{S,T}=\sqrt{\frac{S}{\pi}},\nonumber \\
&&\left(\frac{\partial \Phi}{\partial Q}\right)_{T,P}=\sqrt{\frac{\pi}{S}}\frac{8PS^2-S+\pi Q^2}{8PS^2-S+3\pi Q^2},\nonumber \\
&&\left(\frac{\partial T}{\partial P}\right)_{\Phi,S}=\sqrt{\frac{2S}{\pi}},\nonumber \\
&&\left(\frac{\partial P}{\partial T}\right)_{Q,S}=\sqrt{\frac{\pi}{2S}},
\end{eqnarray}
we then obtain
\begin{eqnarray}
C_{P,Q}&&=C_{P,\Phi}\frac{8PS^2-S+\pi Q^2}{8PS^2-S+3\pi Q^2},
\end{eqnarray}
which is the relation of specific heat capacities of RN-AdS black hole
between grand canonical and canonical ensembles.

We can also get other relations of thermodynamical coefficients for RN-AdS black hole between the two ensembles by using the thermodynamic identity. For example, we can get
\begin{eqnarray}
\alpha_{P,Q}=\alpha_{P,\Phi}\frac{8PS^2-S+\pi Q^2}{8PS^2-S+3\pi Q^2},\nonumber \\
\kappa_{T,Q}=\kappa_{T,\Phi}\frac{8PS^2-S+\pi Q^2}{8PS^2-S+3\pi Q^2},
\end{eqnarray}
where $\alpha_{P,Q}$ and $\alpha_{P,\Phi}$ denotes volume expansion coefficient in the canonical and grand canonical ensembles respectively,
$\kappa_{T,Q}$ and $\kappa_{T,\Phi}$ denotes isothermal compressibility coefficient in the canonical and grand canonical ensembles respectively.

\section{New Maxwell' s law in RN-AdS black hole}

In an usual thermodynamic system with two pairs of intensive/extensive variables $(T,S)$ and $(P,V)$, for the first order phase transition we have usual Maxwell' s equal area law
\begin{eqnarray}
&&\oint VdP=0,\label{VdP}
\end{eqnarray}
in which the ``oscillating" part (theoretical curve) of the isotherm $T<T_c$ can be replaced by an isobar(actual measurement) in $P-V$ plane.

However, in the RN-AdS black hole the actual evolution curve will be complicated and may be not a simple straight line. In general, we cannot use the usual Maxwell' s equal area law to ensure the theoretical curves agree with actual evolution curve for the first order phase transition. Therefore, we should find corresponding new law.

Starting from the Stokes theorem, for the first order phase transition in the RN-AdS black hole, we obtain
\begin{eqnarray}
&&\oint \bigg\{-SdT + \Phi dQ + VdP \bigg\} \nonumber \\
&&=\int\int \bigg\{\left[ \left(\frac{\partial V}{\partial Q}\right)_{P,T}-\left(\frac{\partial \Phi}{\partial P}\right)_{T,Q}\right]dQdP \nonumber \\
&&+\left[ \left(\frac{\partial S}{\partial P}\right)_{T,Q}+\left(\frac{\partial V}{\partial T}\right)_{P,Q}\right]dPdT\nonumber \\
&&+\left[\left(\frac{\partial S}{\partial Q}\right)_{P,T}+\left(\frac{\partial \Phi}{\partial T}\right)_{P,Q} \right]dTdQ \bigg\}=0,\label{Tf}
\end{eqnarray}
where we used relations $\left(\frac{\partial V}{\partial Q}\right)_{P,T}=\left(\frac{\partial \Phi}{\partial P}\right)_{T,Q}$, $\left(\frac{\partial S}{\partial P}\right)_{T,Q}=-\left(\frac{\partial V}{\partial T}\right)_{P,Q}$ and $\left(\frac{\partial S}{\partial Q}\right)_{P,T}=-\left(\frac{\partial \Phi}{\partial T}\right)_{P,Q}$. The ``P-V-Q" diagram is depicted in Fig \ref{PVQ}.

\FIGURE{
\includegraphics[scale=0.8]{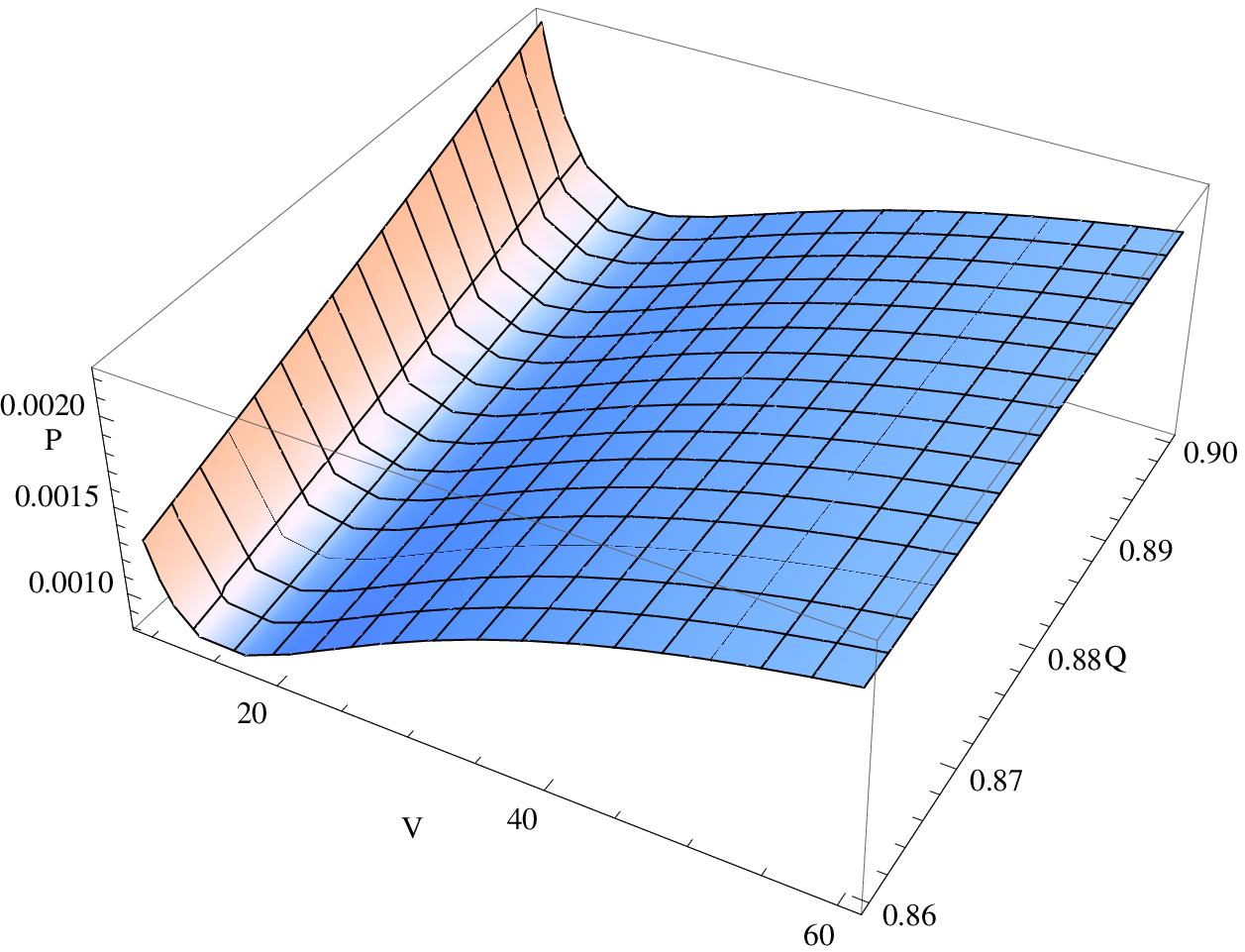}
\caption{\label{PVQ} P-V-Q diagram of the RN-AdS black hole for a fixed T.  }
}

If we only fixed $T<T_c$ as usual, we note that  Eq.(\ref{Tf}) becomes
\begin{eqnarray}
\oint  \bigg\{\Phi dQ + VdP \bigg\} =0,\label{Tf1}
\end{eqnarray}
which is different from the usual Maxwell' s equal area law (\ref{VdP}). We may call it new Maxwell's law which tell us that the evolution curve is not only rely on the $V$ and $P$, but also on the $\Phi$ and $Q$.

Of course, for a thermodynamic system with two pairs of intensive/extensive variables, say  $(T,S)$ and $(P,V)$,   Eq.(\ref{Tf}) reduces to the usual Maxwell' equal area law.

\section{Conclusions}

The thermodynamic behaviors of black holes have been studied for many years and it was expected that usual thermodynamic laws can be extended to black hole systems.  In this paper we studied the thermodynamic properties of RN-AdS black hole in full phase space. We found a generalized thermodynamic identity and then we can easily obtained the relations of thermodynamical coefficients for the RN-AdS black hole between the canonical and grand canonical ensembles. Furthermore, we showed that the usual Maxwell' s equal area law must be extended to a new form in the RN-AdS black hole. The new Maxwell's law tell us how the complicated theoretical curves agree  with a actual evolution in the RN-AdS black hole.

\begin{acknowledgments}
This work is supported by the  National Natural Science Foundation of China under Grant Nos. 11475061; the Open Project Program of State Key Laboratory of Theoretical Physics, Institute of Theoretical Physics, Chinese Academy of Sciences, China (No.Y5KF161CJ1).
\end{acknowledgments}

\end{document}